\documentclass[10pt,twocolumn,twoside]{IEEEtran}
\usepackage{graphicx}
\usepackage{amsmath}
\usepackage{array}
\newcommand{\PreserveBackslash}[1]{\let\temp=\\#1\let\\=\temp}
\newcolumntype{C}[1]{>{\PreserveBackslash\centering}p{#1}}
\newcolumntype{R}[1]{>{\PreserveBackslash\raggedleft}p{#1}}
\newcolumntype{L}[1]{>{\PreserveBackslash\raggedright}p{#1}}
\usepackage[usenames]{color}
\usepackage{colortbl,booktabs}
\usepackage{tabularx}
\usepackage{multicol}
\usepackage{booktabs}
\usepackage[Symbol]{upgreek}
\usepackage{subfigure}
\usepackage{bm}
\usepackage{cite}
\usepackage{balance}
\usepackage[boxed]{algorithm2e}

\usepackage{threeparttable}
\usepackage{amsmath}
\usepackage{dcolumn}
\usepackage{multirow}
\usepackage{booktabs}
\usepackage{amsfonts}
\newcolumntype{d}[1]{D{.}{.}{#1}}

\def \qed {\hfill \vrule height6pt width 6pt depth 0pt}

\begin{document}

\bibliographystyle{IEEEtran} 
\title{Energy-Efficient Hybrid Analog and Digital Precoding for mmWave MIMO Systems with Large Antenna Arrays}

\author{Xinyu Gao,~\IEEEmembership{Student Member,~IEEE}, Linglong Dai,~\IEEEmembership{Senior Member,~IEEE}, Shuangfeng Han,~\IEEEmembership{Member,~IEEE},  \\ Chih-Lin I,~\IEEEmembership{Senior Member,~IEEE}, and Robert W. Heath Jr.,~\IEEEmembership{Fellow,~IEEE}

\thanks{Part of this work has been accepted by IEEE International Conference on Communications (ICC), London, UK, June, 2015.}
\thanks{X. Gao and L. Dai are with the Tsinghua National Laboratory
for Information Science and Technology (TNList), Department of Electronic Engineering, Beijing 100084, China (e-mail: daill@tsinghua.edu.cn).}
\thanks{S. Han and C. I are with the Green Communication Research Center, China Mobile Research
Institute, Beijing 100053, China (e-mail: hanshuangfeng@chinamobile.com).}
\thanks{R. Heath is with The University of Texas at Austin, Austin, TX 78712, USA (e-mail: rheath@utexas.edu).}
\thanks{This work was supported by National Key Basic Research Program of
China (Grant No. 2013CB329203), National 863 Program (Grant No. 2014AA01A704), and National Natural Science Foundation
of China (Grant No. 61271266).}}

\maketitle
\begin{abstract}
Millimeter wave (mmWave) MIMO will likely use hybrid analog and digital precoding, which uses a small number of RF chains to avoid energy consumption associated with mixed signal components like analog-to-digital components not to mention baseband processing complexity. However, most hybrid precoding techniques consider a fully-connected architecture requiring a large number of phase shifters, which is also energy-intensive. In this paper, we focus on the more energy-efficient hybrid precoding with sub-connected architecture, and propose a successive interference cancelation (SIC)-based hybrid precoding with near-optimal performance and low complexity. Inspired by the idea of SIC for multi-user signal detection, we first propose to decompose the total achievable rate optimization problem with non-convex constraints into a series of simple sub-rate optimization problems, each of which only considers one sub-antenna array. Then, we prove that maximizing the achievable sub-rate of each sub-antenna array is equivalent to simply seeking a precoding vector sufficiently close (in terms of Euclidean distance) to the unconstrained optimal solution. Finally, we propose a low-complexity algorithm to realize SIC-based hybrid precoding, which can avoid the need for the singular value decomposition (SVD) and matrix inversion. Complexity evaluation shows that the complexity of SIC-based hybrid precoding is only about 10\% as complex as that of the recently proposed spatially sparse precoding in typical mmWave MIMO systems. Simulation results verify the near-optimal performance of SIC-based hybrid precoding.
\end{abstract}


\section{Introduction}\label{S1}

\IEEEPARstart The integration of  millimeter-wave (mmWave) and multiple-input multiple-output (MIMO) can achieve orders of magnitude increase in rates due to larger bandwidth and greater spectral efficiency~\cite{bai2014coverage}. This makes mmWave MIMO as a promising technique for future 5G wireless communication systems~\cite{pi2011introduction}. On one hand, the decreased wavelength associated with high frequencies of mmWave enables a large antenna array to be packed in small physical dimension~\cite{han2015large}. On the other hand, the large antenna array can provide sufficient antenna gain to compensate for the severe attenuation of mmWave signals due to path loss, oxygen absorption, and rainfall effect~\cite{wei2014key}. Additionally, the large antenna array can also support the transmission of multiple data streams to improve the spectral efficiency through the use of precoding~\cite{alkhateeb2014mimo}.

For MIMO in conventional cellular frequency band (e.g., 2-3 GHz), precoding is entirely realized in the digital domain to cancel the interferences between different data streams. Digital precoding requires an expensive radio frequency (RF) chain (including digital-to-analog converter, up converter, etc.) for every  antenna. In mmWave MIMO system with a large number of antennas, it will bring prohibitively high energy consumption and hardware complexity. To solve this problem, mmWave MIMO prefers the more energy-efficient hybrid analog and digital precoding~\cite{han2015large}, which can significantly reduce the number of required RF chains. Specifically, the transmitted signals are first precoded by the digital precoding of a small dimension to guarantee the performance, and then precoded again by the analog precoding of a large dimension to save the energy consumption and reduce the hardware complexity.

To realize the hybrid precoding in practice, two categories of techniques have been proposed recently. The first category is based on the spatially sparse precoding~\cite{el2013spatially,lee2014hybrid,chen2015an}, which formulates the achievable rate optimization problem as a sparse approximation problem and solves it by orthogonal matching pursuit (OMP)~\cite{tropp2007signal} to achieve the near-optimal performance. The second category of hybrid precoding based on codebook is proposed in~\cite{roh2014millimeter,kim2013tens,kim2013low}, which involves an iterative searching procedure among the predefined codebook to find the optimal hybrid precoding matrix. These algorithms are all designed for the hybrid precoding with fully-connected architecture, where each RF chain is connected to all BS antennas via phase shifters. As the number of BS antennas is very large (e.g., 256 as considered in~\cite{el2013spatially}), the fully-connected architecture has two possible limitations. First, it requires thousands of phase shifters like the giant phased array radar to realize the analog precoding~\cite{balanis2012antenna}, which leads to both high energy consumption and hardware complexity. Second, each RF chain will drive hundreds of BS antennas, which is also energy-intensive~\cite{balanis2012antenna}. By contrast, the hybrid precoding with sub-connected architecture, where each RF chain is connected to only a subset of BS antennas, can reduce the number of required phase shifters without obvious performance loss~\cite{han2015large}. Therefore, the sub-connected architecture is expected to be more energy-efficient and implementation-practical for mmWave MIMO systems. Unfortunately, the design of hybrid precoding with sub-connected architecture is still an open problem~\cite{han2015large,han2014reference}, as the sub-connected architecture changes the constraints on the original hybrid precoding problem.

In this paper, we propose a successive interference cancelation (SIC)-based hybrid precoding with sub-connected architecture. The contributions of this paper can be summarized as follows.

1) Inspired by the idea of SIC derived for multi-user signal detection~\cite{liang2008relationship}, we propose to decompose the total achievable rate optimization problem with non-convex constraints into a series of simple sub-rate optimization problems, each of which only considers one sub-antenna array. Then, we can maximize the achievable sub-rate of each sub-antenna array one by one until the last sub-antenna array is considered.

2) We prove that maximizing the achievable sub-rate of each sub-antenna array is equivalent to seeking a precoding vector which has the smallest Euclidean distance to the unconstrained optimal solution. Based on this fact, we can easily obtain the optimal precoding vector for each sub-antenna array.

3) We further propose a low-complexity algorithm to realize the SIC-based precoding, which avoids the need for singular value decomposition (SVD) and matrix inversion. Complexity evaluation shows that the complexity of SIC-based precoding is only about 10\% as complex as that of the spatially sparse precoding~\cite{el2013spatially} in typical mmWave MIMO systems, while it can still achieve the near-optimal performance as verified by simulation results.

It is worth pointing out that to the best of the authors' knowledge, our work in this paper is the first one that considers the hybrid precoding design with sub-connected architecture.

The rest of the paper is organized as follows. Section~\ref{S2} briefly introduces the system model of mmWave MIMO. Section~\ref{S3} specifies the proposed SIC-based hybrid precoding, together with the complexity evaluation. The simulation results of the achievable rate are shown in Section~\ref{S4}. Finally, conclusions are drawn in Section~\ref{S5}.

{\it Notation}: Lower-case and upper-case boldface letters denote vectors and matrices, respectively;  ${( \cdot )^T}$, ${( \cdot )^H}$, ${( \cdot )^{ - 1}}$, and ${\left|  \cdot  \right|}$  denote the transpose, conjugate transpose, inversion, and determinant of a matrix, respectively; ${{\left\|  \cdot  \right\|_1}}$ and ${{\left\|  \cdot  \right\|_2}}$  denote the ${{l_1}}$- and ${{l_2}}$-norm of a vector, respectively; ${{\left\|  \cdot  \right\|_F}}$  denotes the Frobenius norm of a matrix; ${{\mathop{\rm Re}\nolimits} \{  \cdot \} }$ and ${{\mathop{\rm Im}\nolimits} \{  \cdot \} }$ denote the real part and imaginary part of a complex number, respectively; ${\mathbb{E}( \cdot )}$  denotes the expectation; Finally, ${{\bf{I}}_N}$ is the  $ N \times N $  identity matrix.

\section{System Model}\label{S2}

\begin{figure}[tp]
\begin{center}
\includegraphics[width=1\linewidth]{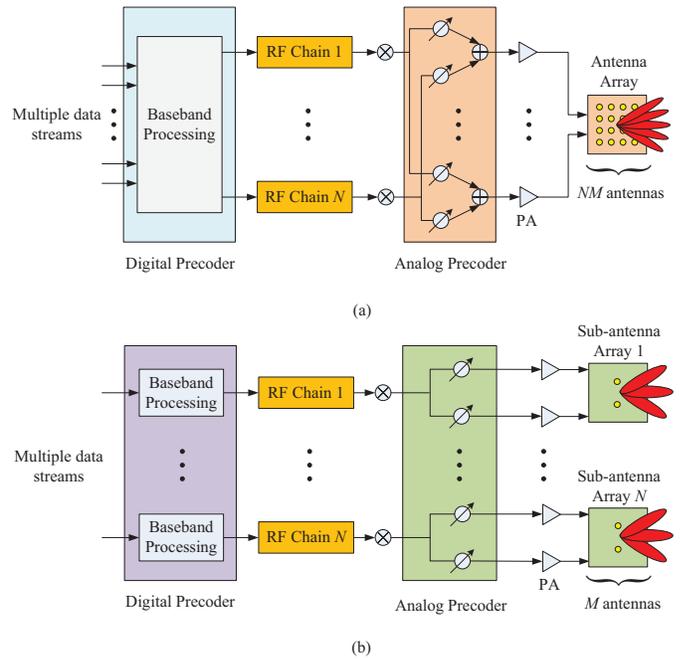}
\end{center}
\vspace*{-4mm}\caption{Two typical architectures of the hybrid precoding in mmWave MIMO systems: (a) Fully-connected architecture, where each RF chain is connected to all BS antennas; (b) Sub-connected architecture, where each RF chain is connected to only a subset of BS antennas.} \label{FIG1}
\vspace*{1mm}
\end{figure}

Fig. 1 illustrates two typical architectures for hybrid precoding in mmWave MIMO systems, i.e., the fully-connected architecture as shown in Fig. 1 (a) and the sub-connected architecture as shown in Fig. 1 (b). In both cases the BS has ${NM}$  antennas but only ${N}$ RF chains. From Fig. 1, we observe that the sub-connected architecture will likely be more energy-efficient, since it only requires ${NM}$ phase shifters, while the fully-connected architecture requires ${N^{2}M}$ phase shifters. To fully achieve the spatial multiplexing gain, the BS usually transmits ${N}$ independent data streams to users employing ${K}$ receive antennas~\cite{han2015large}.

In the sub-connected architecture as shown in Fig. 1 (b), ${N}$ data streams in the baseband are precoded by the digital precoder ${\bf{D}}$. In cases where complexity is a concern, ${\bf{D}}$ can be further specialized to be a diagonal matrix as ${{\bf{D}} = {\rm{diag}}\left[ {{d_1},{d_2}, \cdots ,{d_N}} \right]}$, where ${{d_n} \in \mathbb{R}}$ for ${n = 1, 2, \cdots ,N}$~\cite{han2015large}. Then the role of ${\bf{D}}$ essentially performs some power allocation. After passing through the corresponding  RF chain, the digital-domain signal from each RF chain is delivered to only ${M}$ phase shifters~\cite{hur2013millimeter} to perform the analog precoding, which can be denoted by the analog weighting vector ${{{\bf{\bar a}}_n} \in {\mathbb{C}^{M \times 1}}}$, whose elements have the same amplitude ${1/\sqrt M }$ but different phases~\cite{hur2013millimeter}.
After the analog precoding, each data stream is finally transmitted by a sub-antenna array with only ${M}$  antennas associated with the corresponding RF chain. Then, the received signal vector ${{\bf{y}} = {[{y_1}, {y_2}, \cdot  \cdot  \cdot ,{y_K}]^T}}$  at the user in a narrowband system \footnote{While mmWave systems are expected to be broadband as in prior work~\cite{pi2011introduction}, the narrowband system can be regarded as a reasonable first step. The extension to broadband system is an interesting topic of future work.} can be presented as
\begin{equation}\label{eq1}
{\bf{y}} = \rho {\bf{HADs}} + {\bf{n}} = \rho {\bf{HPs}} + {\bf{n}},
\end{equation}
where ${\rho }$ is the average received power; ${{\bf{H}} \in \mathbb{C}{^{K \times NM}}}$ denotes the channel matrix, ${{\bf{A}}}$ is the ${NM \times N}$ analog precoding matrix comprising ${N}$ analog weighting vectors ${\left\{ {{{{\bf{\bar a}}}_m}} \right\}_{m = 1}^N}$ as
\begin{equation}\label{eq2}
{\bf{A}} = {\left[ {\begin{array}{*{20}{c}}
{{{{\bf{\bar a}}}_1}}&{\bf{0}}& \ldots &{\bf{0}}\\
{\bf{0}}&{{{{\bf{\bar a}}}_2}}&{}&{\bf{0}}\\
 \vdots &{}& \ddots & \vdots \\
{\bf{0}}&{\bf{0}}& \ldots &{{{{\bf{\bar a}}}_N}}
\end{array}} \right]_{NM \times N}},
\end{equation}
${{\bf{s}} = {[{s_1}, {s_2}, \cdot  \cdot  \cdot ,{s_N}]^T}}$  represents the transmitted signal vector in the baseband, and usually ${\mathbb{E}({\bf{s}}{{\bf{s}}^H}) = \frac{1}{N}{{\bf{I}}_N}}$ is assumed for the normalized signal power~\cite{el2013spatially}. ${{\bf{P}} = {\bf{AD}}}$ presents the hybrid precoding matrix of size ${NM \times N}$, which satisfies ${{\left\| {\bf{P}} \right\|_F} \le N}$ to meet the total transmit power constraint~\cite{el2013spatially}. Finally, ${{\bf{n}} = {[{n_1}, {n_2}, \cdot  \cdot  \cdot ,{n_N}]^T}}$  is an additive white Gaussian noise (AWGN) vector, whose entries follow the independent and identical distribution (i.i.d.) ${{\cal C}{\cal N}(0,{\sigma ^2})}$.

It is known that mmWave channel ${\bf{H}}$ will not likely follow the rich-scattering model assumed at low frequencies due to the limited number of scatters in the mmWave prorogation environment~\cite{pi2011introduction}. In this paper, we adopt  the geometric Saleh-Valenzuela channel model for mmWave communications, which was also used in related work in~\cite{alkhateeb2014channel} as
\begin{equation}\label{eq3}
{\bf{H}} = \gamma \sum\limits_{l = 1}^L {{\alpha _l}{\Lambda _r}} \left( {\phi _l^r,\theta _l^r} \right){\Lambda _t}\left( {\phi _l^t,\theta _l^t} \right){{\bf{f}}_r}\left( {\phi _l^r,\theta _l^r} \right){\bf{f}}_t^H\left( {\phi _l^t,\theta _l^t} \right),
\end{equation}
where ${\gamma  = \sqrt {\frac{{NMK}}{L}} }$ is a normalization factor, ${L}$  is the number of effective  channel paths corresponding to the limited number of scatters, and we usually have ${L \le N}$ for  mmWave communication systems. ${{\alpha _l} \in \mathbb{C}}$ is the gain of the  ${l}$th path. ${\phi _l^t}$ (${{\theta _l^t}}$) and ${\phi _l^r}$ (${{\theta _l^r}}$) are the azimuth (elevation) angles of departure and arrival (AoDs/AoAs), respectively. ${{\Lambda _t}\left( {\phi _l^t, \theta _l^t } \right)}$  and ${{\Lambda _r}\left( {\phi _l^r, \theta_l^r} \right)}$  denote the transmit and receive antenna array gain at a specific AoD and AoA, respectively. For simplicity but without loss of generality, ${{\Lambda _t}\left( {\phi _l^t, \theta _l^t } \right)}$ and ${{\Lambda _r}\left( {\phi _l^r, \theta_l^r} \right)}$  can be set as one within the range of AoDs/AoAs~\cite{alkhateeb2013hybrid}. Finally, ${{{\bf{f}}_t}\left( {\phi _l^t, \theta_l^t} \right)}$  and  ${{{\bf{f}}_r}\left( {\phi _l^r, \theta_l^r} \right)}$ are the antenna array response vectors  depending on the antenna array structures at the BS and the user, respectively. For the uniform linear array (ULA) with ${U}$ elements, the array response vector can be presented as~\cite{balanis2012antenna}
\begin{equation}\label{eq4}
{{\bf{f}}_{{\rm{ULA}}}}\left( \phi  \right) = \frac{1}{{\sqrt U }}{\left[ {1,{e^{j\frac{{2\pi }}{\lambda }d\sin \left( \phi  \right)}}, \cdot  \cdot  \cdot ,{e^{j(U - 1)\frac{{2\pi }}{\lambda }d\sin \left( \phi  \right)}}} \right]^T},
\end{equation}
where ${\lambda }$ denotes the wavelength of the signal, and ${d}$ is the antenna spacing. Note that here we abandon the subscripts ${\left\{ {t,r} \right\}}$ in~(\ref{eq3}) and we also do not include ${\theta}$ in~(\ref{eq4}) since the ULA response vector is independent of the elevation angle. Additionally, when we consider the uniform planar array (UPA) with ${W_1}$ and ${W_2}$ elements (${{W_1}{W_2}=U}$) on horizon and vertical, respectively, the array response vector can be given by~\cite{balanis2012antenna}
\begin{align}\label{eq5}
{{\bf{f}}_{{\rm{UPA}}}}\left( {\phi ,\theta } \right) &= \frac{1}{{\sqrt U }}\left[ {1, \cdots ,{e^{j\frac{{2\pi }}{\lambda }d\left( {x\sin \left( \phi  \right)\sin \left( \theta  \right) + y\cos \left( \theta  \right)} \right)}},} \right. \nonumber \\
&{\left. { \cdot  \cdot  \cdot ,{e^{j\frac{{2\pi }}{\lambda }d\left( {\left( {{W_1} - 1} \right)\sin \left( \phi  \right)\sin \left( \theta  \right) + \left( {{W_2} - 1} \right)\cos \left( \theta  \right)} \right)}}} \right]^T},
\end{align}
where ${0 \le x \le \left( {{W_1} - 1} \right)}$ and ${0 \le y \le \left( {{W_2} - 1} \right)}$.

\vspace*{+2mm}
\section{SIC-Based Hybrid Precoding for mmWave MIMO Systems}\label{S3}
In this section, we propose a low-complexity SIC-based hybrid precoding to achieve the near-optimal performance. The evaluation of computational complexity is also provided to show its advantages over current solutions.
\subsection{Structure of SIC-based hybrid precoding}\label{S3.1}
The final aim of precoding is to maximize  the total achievable rate ${R}$ of mmWave MIMO systems, which can be expressed as~\cite{el2013spatially}
\begin{equation}\label{eq6}
R = {\log _2}\left( {\left| {{{\bf{I}}_N} + \frac{\rho }{{N{\sigma ^2}}}{\bf{HP}}{{\bf{P}}^H}{{\bf{H}}^H}} \right|} \right).
\end{equation}
According to the system model (1) in Section II, since the hybrid precoding matrix ${\bf{P}}$ can be represented as ${{\bf{P}} = {\bf{AD}} = {\rm{diag}}\left\{ {{{{\bf{\bar a}}}_1}, \cdots ,{{{\bf{\bar a}}}_N}} \right\} \cdot {\rm{diag}}\left\{ {{d_1}, \cdots ,{d_N}} \right\}}$, there are three constraints for the design of ${{\bf{P}}}$.

\vspace*{+2mm}
\noindent\emph{Constraint 1}: ${{\bf{P}}}$ should be a block diagonal matrix similar to the form of ${{\bf{A}}}$ as shown in (2), i.e., ${{\bf{P}} = {\rm{diag}}\left\{ {{{{\bf{\bar p}}}_1}, \cdots ,{{{\bf{\bar p}}}_N}} \right\}}$, where ${{{\bf{\bar p}}_n} = {d_n}{{\bf{\bar a}}_n}}$ is the ${M \times 1}$ non-zero vector of the ${n}$th column ${{{\bf{p}}_n}}$ of ${{\bf{P}}}$, i.e., ${{{\bf{p}}_n} = {\left[ {{{\bf{0}}_{1 \times M\left( {n - 1} \right)}},\;{{{\bf{\bar p}}}_n},\;{{\bf{0}}_{1 \times M\left( {N - n} \right)}}} \right]^T}}$;

\vspace*{+2mm}
\noindent\emph{Constraint 2}: The non-zero elements of each column of ${{\bf{P}}}$ should have the same amplitude, since the digital precoding matrix ${{\bf{D}}}$ is a diagonal matrix, and the amplitude of non-zero elements of the analog precoding matrix ${{\bf{A}}}$ is fixed to ${1/\sqrt M }$;

\vspace*{+2mm}
\noindent\emph{Constraint 3}:  The Frobenius norm of ${{\bf{P}}}$ should satisfy ${{\left\| {\bf{P}} \right\|_F} \le N}$ to meet the total transmit power constraint.

\vspace*{+2mm}
Unfortunately, these non-convex constraints on ${{\bf{P}}}$  make maximizing the  total achievable rate~(\ref{eq6}) very difficult to be solved. However, based on the special block diagonal structure of the hybrid precoding matrix ${{\bf{P}}}$, we can observe that the precoding on different sub-antenna arrays are independent. This inspires us to decompose the total achievable rate~(\ref{eq6}) into a series of sub-rate optimization problems, each of which only considers one sub-antenna array.

In particular, we can divide the hybrid precoding matrix ${{\bf{P}}}$ as ${{\bf{P}} = \left[ {{{\bf{P}}_{N - 1}}\;{{\bf{p}}_N}} \right]}$, where ${{{{\bf{p}}_N}}}$ is the ${N}$th column of ${{\bf{P}}}$, and ${{\bf{P}}_{N - 1}}$ is an ${NM \times (N-1)}$ matrix containing the first (${N-1}$) columns of ${{\bf{P}}}$. Then, the  total achievable rate ${R}$ in~(\ref{eq6}) can be rewritten as
\begin{align}\label{eq7}
R &= {\log _2}\left( {\left| {{{\bf{I}}_N} + \frac{\rho }{{N{\sigma ^2}}}{\bf{HP}}{{\bf{P}}^H}{{\bf{H}}^H}} \right|} \right) \nonumber \\
&= {\log _2}\left( {\left| {{{\bf{I}}_N} + \frac{\rho }{{N{\sigma ^2}}}{\bf{H}}\left[ {{{\bf{P}}_{N - 1}}\;{{\bf{p}}_N}} \right]{{\left[ {{{\bf{P}}_{N - 1}}\;{{\bf{p}}_N}} \right]}^H}{{\bf{H}}^H}} \right|} \right) \nonumber \\
&= {\log _2}\left( {\left| {{{\bf{I}}_N} + \frac{\rho }{{N{\sigma ^2}}}{\bf{H}}{{\bf{P}}_{N - 1}}{\bf{P}}_{N - 1}^H{{\bf{H}}^H}} \right.} \right. \nonumber \\
& \left. {\left. {\quad \quad \quad \quad  + \frac{\rho }{{N{\sigma ^2}}}{\bf{H}}{{\bf{p}}_N}{\bf{p}}_N^H{{\bf{H}}^H}} \right|} \right)
 \nonumber \\
&\mathop  = \limits^{\left( a \right)} {\log _2}\left( {\left| {{{\bf{T}}_{N - 1}}} \right|} \right) + {\log _2}\left( {\left| {{{\bf{I}}_N} + \frac{\rho }{{N{\sigma ^2}}}{\bf{T}}_{N - 1}^{ - 1}{\bf{H}}{{\bf{p}}_N}{\bf{p}}_N^H{{\bf{H}}^H}} \right|} \right) \nonumber \\
&\mathop  = \limits^{\left( b \right)} {\log _2}\left( {\left| {{{\bf{T}}_{N - 1}}} \right|} \right) + {\log _2}\left( {1\! +\! \frac{\rho }{{N{\sigma ^2}}}{\bf{p}}_N^H{{\bf{H}}^H}{\bf{T}}_{N - 1}^{ - 1}{\bf{H}}{{\bf{p}}_N}} \right),
\end{align}
where ${\left( a \right)}$  is obtained by defining the auxiliary matrix ${{{\bf{T}}_{N - 1}} = {{\bf{I}}_N} + \frac{\rho }{{N{\sigma ^2}}}{\bf{H}}{{\bf{P}}_{N - 1}}{\bf{P}}_{N - 1}^H{{\bf{H}}^H}}$, and  ${\left( b \right)}$ is true due to the fact that ${\left| {{\bf{I}} + {\bf{XY}}} \right| = \left| {{\bf{I}}+ {\bf{YX}}} \right|}$ by defining ${{\bf{X}} = {\bf{T}}_{N - 1}^{ - 1}{\bf{H}}{{\bf{p}}_N}}$ and ${{\bf{Y}} = {\bf{p}}_N^H{{\bf{H}}^H}}$.
Note that the second term ${{\log _2}\left( {1 + \frac{\rho }{{N{\sigma ^2}}}{\bf{p}}_N^H{{\bf{H}}^H}{\bf{T}}_{N - 1}^{ - 1}{\bf{H}}{{\bf{p}}_N}} \right)}$ on the right side  of~(\ref{eq7}) is the achievable sub-rate of the ${N}$th sub-antenna array, while the first term ${{\log _2}\left( {\left| {{{\bf{T}}_{N - 1}}} \right|} \right)}$ shares the same form as~(\ref{eq6}). This observation implies that we can further decompose ${{\log _2}\left( {\left| {{{\bf{T}}_{N - 1}}} \right|} \right)}$ using the similar method in~(\ref{eq7}) as
\begin{equation}\label{eq60}
{\log _2}\left( {\left| {{{\bf{T}}_{N - 2}}} \right|} \right) \!+\! {\log _2}\left( {1\! +\! \frac{\rho }{{N{\sigma ^2}}}{\bf{p}}_{N - 1}^H{{\bf{H}}^H}{\bf{T}}_{N - 2}^{ - 1}{\bf{H}}{{\bf{p}}_{N - 1}}} \right) \nonumber.
\end{equation}

Then, after ${N}$ such decompositions, the total achievable rate ${R}$ in~(\ref{eq6}) can be presented as
\begin{align}\label{eq8}
R = \sum\limits_{n = 1}^N {{{\log }_2}\left( {1 + \frac{\rho }{{N{\sigma ^2}}}{\bf{p}}_n^H{{\bf{H}}^H}{\bf{T}}_{n - 1}^{ - 1}{\bf{H}}{{\bf{p}}_n}} \right)},
\end{align}
where we have ${{{\bf{T}}_n}={{\bf{I}}_N} + \frac{\rho }{{N{\sigma ^2}}}{\bf{H}}{{\bf{P}}_n}{\bf{P}}_n^H{{\bf{H}}^H}}$ and ${{{\bf{T}}_0} = {{\bf{I}}_N}}$. From~(\ref{eq8}), we observe that the total achievable rate optimization problem can be transformed into a series of sub-rate optimization problems of sub-antenna arrays, which can be optimized one by one. After that, inspired by the idea of SIC for multi-user signal detection~\cite{liang2008relationship}, we can optimize the achievable sub-rate of the first sub-antenna array and update the matrix ${{{\bf{T}}_1}}$. Then, the similar method can be utilized to optimize the achievable sub-rate of the second sub-antenna array. Such procedure will be executed until the last sub-antenna array is considered. Fig. 2 shows the diagram of the proposed SIC-based hybrid precoding. Next, we will discuss how to optimize the achievable sub-rate of each sub-antenna array.

\begin{figure}[h]
\setlength{\abovecaptionskip}{0pt}
\setlength{\belowcaptionskip}{0pt}
\begin{center}
\hspace*{0mm}\includegraphics[width=1\linewidth]{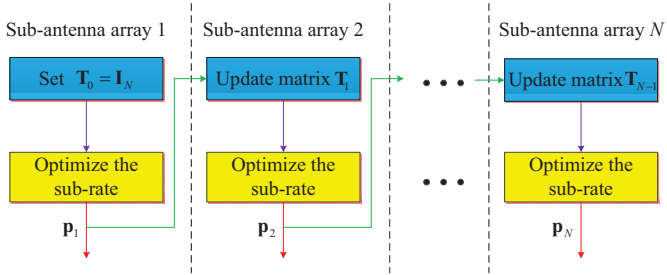}
\end{center}
\caption{Diagram of the proposed SIC-based hybrid precoding.} \label{FIG2}
\end{figure}

\vspace*{-5mm}
\subsection{Solution to the sub-rate optimization problem}\label{S3.2}
In this subsection, we focus on the sub-rate optimization problem of the ${n}$th sub-antenna array, which can be directly applied to other sub-antenna arrays. According to~(\ref{eq8}), the sub-rate optimization problem of the ${n}$th sub-antenna array by designing the ${n}$th precoding vector ${{{{\bf{p}}_n}}}$ can be stated as
\begin{equation}\label{eq9}
{\bf{p}}_n^{{\rm{opt}}} = \mathop {\arg \max }\limits_{{{\bf{p}}_n} \in {\cal F}} {\log _2}\left( {1 + \frac{\rho }{{N{\sigma ^2}}}{\bf{p}}_n^H{{\bf{G}}_{n - 1}}{{\bf{p}}_n}} \right),
\end{equation}
where ${{{\bf{G}}_{n - 1}}}$ is defined as ${{{\bf{G}}_{n - 1}}={{\bf{H}}^H}{\bf{T}}_{n - 1}^{ - 1}{\bf{H}}}$, ${{\cal F}}$ is the set of all feasible vectors satisfying the three constraints described in Section III-A. Note that the ${n}$th precoding vector ${{{{\bf{p}}_n}}}$ only has ${M}$ non-zero elements from the (${M(n-1)+1}$)th one to the (${Mn}$)th one. Therefore, the sub-rate optimization problem~(\ref{eq9}) can be equivalently written as
\begin{equation}\label{eq10}
{\bf{\bar p}}_n^{{\rm{opt}}} = \mathop {\arg \max }\limits_{{{{\bf{\bar p}}}_n} \in \bar {\cal F}} {\log _2}\left( {1 + \frac{\rho }{{N{\sigma ^2}}}{\bf{\bar p}}_n^H{{{\bf{\bar G}}}_{n - 1}}{{{\bf{\bar p}}}_n}} \right),
\end{equation}
where ${{\bar {\cal F}}}$ includes all possible ${M \times 1}$ vectors satisfying \emph{Constraint 2} and \emph{Constraint 3}, ${{{{{\bf{\bar G}}}_{n - 1}}}}$ of size ${M \times M}$ is the corresponding sub-matrix of ${{{{\bf{G}}_{n - 1}}}}$ by only keeping the rows and columns of ${{{{\bf{G}}_{n - 1}}}}$ from the (${M(n-1)+1}$)th one to the (${Mn}$)th one, which can be presented as
\begin{equation}\label{eq11}
{{\bf{\bar G}}_{n-1}} = {\bf{R}}{{\bf{G}}_{n - 1}}{{\bf{R}}^H}= {\bf{R}}{{\bf{H}}^H}{\bf{T}}_{n - 1}^{ - 1}{\bf{H}}{{\bf{R}}^H},
\end{equation}
where ${{\bf{R}} = \left[ {\begin{array}{*{20}{c}}
{{{\bf{0}}_{M \times M\left( {n - 1} \right)}}}&{{{\bf{I}}_M}}&{{{\bf{0}}_{M \times M\left( {N - n} \right)}}}
\end{array}} \right]}$ is the corresponding selection matrix.

Define the singular value decomposition (SVD) of the Hermitian matrix ${{{{{\bf{\bar G}}}_{n - 1}}}}$ as ${{{\bf{\bar G}}_{n - 1}} = {\bf{V\Sigma }}{{\bf{V}}^H}}$, where ${{\bf{\Sigma }}}$ is an ${M \times M}$  diagonal matrix containing the singular values of ${{{{{\bf{\bar G}}}_{n - 1}}}}$ in a decreasing order, and ${{\bf{V}}}$ is an ${M \times M}$ unitary matrix. It is known that the optimal unconstrained precoding vector of~(\ref{eq10}) is the first column ${{{\bf{v}}_1}}$ of ${{\bf{V}}}$,  i.e., the first right singular vector of ${{{{{\bf{\bar G}}}_{n - 1}}}}$~\cite{el2013spatially}. According to the constraints mentioned in Section~\ref{S3}-A, we cannot directly choose ${{\bf{\bar p}}_n^{{\rm{opt}}}}$ as ${{{\bf{v}}_1}}$  since the elements of ${{{\bf{v}}_1}}$ do not obey the constraint of same amplitude (i.e., \emph{Constraint 2}). To find a practical solution to the sub-rate optimization problem~(\ref{eq10}), we need to further convert~(\ref{eq10}) into another form, which is given by the following \textbf{Proposition 1}.

\vspace*{+2mm} \noindent\textbf{Proposition 1}. {\it The optimization problem~(\ref{eq10})
\begin{equation}\label{eq61}
{\bf{\bar p}}_n^{{\rm{opt}}} = \mathop {\arg \max }\limits_{{{{\bf{\bar p}}}_n} \in \bar {\cal F}} {\log _2}\left( {1 + \frac{\rho }{{N{\sigma ^2}}}{\bf{\bar p}}_n^H{{{\bf{\bar G}}}_{n - 1}}{{{\bf{\bar p}}}_n}} \right) \nonumber
\end{equation}
is equivalent to the following problem
\begin{equation}\label{eq12}
{\bf{\bar p}}_n^{{\rm{opt}}} = \mathop {\arg \min }\limits_{{{{\bf{\bar p}}}_n} \in \bar {\cal F}} \left\| {{{\bf{v}}_1} - {{{\bf{\bar p}}}_n}} \right\|_2^2,
\end{equation}}
where ${{{\bf{v}}_1}}$ is the first right singular vector of ${{{{{\bf{\bar G}}}_{n - 1}}}}$.

\vspace*{+2mm}
\textit{Proof:} See Appendix A. \qed

\textbf{Proposition 1} indicates that we can find a feasible precoding vector ${{{{{\bf{\bar p}}}_n}}}$, which is sufficiently close (in terms of Euclidean distance) to the optimal but unpractical precoding vector ${{{\bf{v}}_1}}$, to maximize the achievable sub-rate of the ${n}$th sub-antenna array. Since ${{{\bf{\bar p}}_n} = {d_n}{{\bf{\bar a}}_n}}$ according to~(\ref{eq1}), the target ${\left\| {{{\bf{v}}_1} - {{{\bf{\bar p}}}_n}} \right\|_2^2}$  in~(\ref{eq12}) can be rewritten as
\begin{align}\label{eq13}
&\left\| {{{\bf{v}}_1} - {{{\bf{\bar p}}}_n}} \right\|_2^2 \nonumber \\
& \quad= {\left( {{{\bf{v}}_1} - {d_n}{{{\bf{\bar a}}}_n}} \right)^H}\left( {{{\bf{v}}_1} - {d_n}{{{\bf{\bar a}}}_n}} \right) \nonumber \\
& \quad= {\bf{v}}_1^H{{\bf{v}}_1} + d_n^2{\bf{\bar a}}_n^H{{{\bf{\bar a}}}_n} - 2{d_n}{\mathop{\rm Re}\nolimits} \left( {{\bf{v}}_1^H{{{\bf{\bar a}}}_n}} \right) \nonumber \\
& \quad \mathop  = \limits^{\left( a \right)} 1 + d_n^2 - 2{d_n}{\mathop{\rm Re}\nolimits} \left( {{\bf{v}}_1^H{{{\bf{\bar a}}}_n}} \right) \nonumber \\
& \quad = {\left( {{d_n} - {\mathop{\rm Re}\nolimits} \left( {{\bf{v}}_1^H{{{\bf{\bar a}}}_n}} \right)} \right)^2} + \left( {1 - {{\left[ {{\mathop{\rm Re}\nolimits} \left( {{\bf{v}}_1^H{{{\bf{\bar a}}}_n}} \right)} \right]}^2}} \right),
\end{align}
where (a) is obtained based on the facts that ${{\bf{v}}_1^H{{\bf{v}}_1} = 1}$ and ${{\bf{\bar a}}_n^H{{\bf{\bar a}}_n} = 1}$, since ${{{\bf{v}}_1}}$ is the first column of the unitary matrix ${{\bf{V}}}$ and each element of ${{{\bf{\bar a}}_n}}$ has the same amplitude ${1/\sqrt M }$.

From~(\ref{eq13}), we observe that the distance between ${{{{{\bf{\bar p}}}_n}}}$ and ${{{\bf{v}}_1}}$ consists of two parts. The first one is ${{\left( {{d_n} - {\mathop{\rm Re}\nolimits} \left( {{\bf{v}}_1^H{{{\bf{\bar a}}}_n}} \right)} \right)^2}}$, which can be minimized to zero by choosing ${{d_n} = {\mathop{\rm Re}\nolimits} \left( {{\bf{v}}_1^H{{{\bf{\bar a}}}_n}} \right)}$. The second one is ${\left( {1 - {{\left[ {{\mathop{\rm Re}\nolimits} \left( {{\bf{v}}_1^H{{{\bf{\bar a}}}_n}} \right)} \right]}^2}} \right)}$, which can be minimized by maximizing ${\left| {{\rm{Re}}\left( {{\bf{v}}_1^H{{{\bf{\bar a}}}_n}} \right)} \right|}$. Note that both ${{{\bf{\bar a}}_n}}$ and ${{{\bf{v}}_1}}$ have a fixed power of one, i.e., ${{\bf{v}}_1^H{{\bf{v}}_1} = 1}$ and ${{\bf{\bar a}}_n^H{{\bf{\bar a}}_n} = 1}$. Therefore, the optimal ${{\bf{\bar a}}_n^{{\rm{opt}}}}$ to maximize ${\left| {{\rm{Re}}\left( {{\bf{v}}_1^H{{{\bf{\bar a}}}_n}} \right)} \right|}$ is
\begin{equation}\label{eq14}
{\bf{\bar a}}_n^{{\rm{opt}}} = \frac{1}{{\sqrt M }}{e^{j{\rm{angle}}({{\bf{v}}_{\rm{1}}})}},
\end{equation}
where ${{\rm{angle(}}{{\bf{v}}_1}{\rm{)}}}$ denotes the phase vector of ${{{\bf{v}}_1}}$, i.e., each element of ${{\bf{\bar a}}_n^{{\rm{opt}}}}$ shares the same phase as the corresponding element of ${{{\bf{v}}_1}}$. Accordingly, the optimal choice of ${d_n^{{\rm{opt}}}}$ is
\begin{equation}\label{eq15}
d_n^{{\rm{opt}}}\! =\!{\mathop{\rm Re}\nolimits} \left( {{\bf{v}}_1^H{{{\bf{\bar a}}}_n}} \right)\! = \!\frac{1}{{\sqrt M }}{\mathop{\rm Re}\nolimits} \left( {{\bf{v}}_1^H{e^{j{\rm{angle}}({{\bf{v}}_{\rm{1}}})}}} \right)\! = \!\frac{{{{\left\| {{{\bf{v}}_1}} \right\|}_1}}}{{\sqrt M }}.
\end{equation}
Based on~(\ref{eq14}) and~(\ref{eq15}), the optimal solution ${{\bf{\bar p}}_n^{{\rm{opt}}}}$ to the optimization problem~(\ref{eq12}) (or equivalently~(\ref{eq10})) can be obtained by
\begin{equation}\label{eq16}
{\bf{\bar p}}_n^{{\rm{opt}}} = d_n^{{\rm{opt}}}{\bf{\bar a}}_n^{{\rm{opt}}} = \frac{1}{M}{\left\| {{{\bf{v}}_1}} \right\|_1}{e^{j{\rm{angle}}({{\bf{v}}_{\rm{1}}})}}.
\end{equation}

It is worth pointing out that ${{{\bf{v}}_1}}$ is the first column of the unitary matrix ${{\bf{V}}}$, each element ${v_i}$ of ${{{\bf{v}}_1}}$ (for ${i = 1, \cdots ,M}$) has the amplitude less than one. Therefore, we have ${\left\| {{\bf{\bar p}}_n^{{\rm{opt}}}} \right\|_2^2 \le 1}$. Note that for all sub-antenna arrays, the optimal solution ${{\bf{\bar p}}_n^{{\rm{opt}}}}$ for ${n = 1, 2, \cdots ,N}$ have a similar form. Thus, we can conclude that
\begin{equation}\label{eq17}
\left\| {{{\bf{P}}^{{\rm{opt}}}}} \right\|_F^2 = \left\| {{\rm{diag}}\left\{ {{\bf{\bar p}}_1^{{\rm{opt}}}, \cdots ,{\bf{\bar p}}_N^{{\rm{opt}}}} \right\}} \right\|_F^2 \le N,
\end{equation}
which demonstrates that the total transmit power constraint  (\emph{Constraint 3}) is satisfied.

After we have acquired ${{\bf{\bar p}}_n^{{\rm{opt}}}}$ for the ${n}$th sub-antenna array, the matrices ${{{\bf{T}}_n}= {{\bf{I}}_N} + \frac{\rho }{{N{\sigma ^2}}}{\bf{H}}{{\bf{P}}_n}{\bf{P}}_n^H{{\bf{H}}^H}}$~(\ref{eq8}) and ${{{\bf{\bar G}}_n} = {\bf{R}}{{\bf{H}}^H}{\bf{T}}_n^{ - 1}{\bf{H}}{{\bf{R}}^H}}$~(\ref{eq11}) can be updated. Then, the method described above for the ${n}$th sub-antenna array can be reused again to optimize the achievable sub-rate of the (${n+1}$)th sub-antenna array. To sum up, solving the sub-rate optimization problem of the ${n}$th sub-antenna array consists of the following three steps.

\vspace*{+2mm}
\noindent\emph{Step 1}:  Execute the SVD of ${{{\bf{\bar G}}_{n - 1}}}$ to obtain ${{{\bf{v}}_1}}$;

\vspace*{+2mm}
\noindent\emph{Step 2}:  Let ${{\bf{\bar p}}_n^{{\rm{opt}}} = \frac{1}{M}{\left\| {{{\bf{v}}_1}} \right\|_1}{e^{j{\rm{angle}}({{\bf{v}}_{\rm{1}}})}}}$ as the optimal solution to the current ${n}$th sub-antenna array;

\vspace*{+2mm}
\noindent\emph{Step 3}:  Update matrices ${{{\bf{T}}_n}= {{\bf{I}}_N} + \frac{\rho }{{N{\sigma ^2}}}{\bf{H}}{{\bf{P}}_n}{\bf{P}}_n^H{{\bf{H}}^H}}$ and ${{{\bf{\bar G}}_n} = {\bf{R}}{{\bf{H}}^H}{\bf{T}}_n^{ - 1}{\bf{H}}{{\bf{R}}^H}}$ for the next ${(n+1)}$th sub-antenna array.

\vspace*{+2mm}
Note that although we can obtain the optimal solution ${{\bf{\bar p}}_n^{{\rm{opt}}}}$ by the method above, we need to compute the SVD of ${{{\bf{\bar G}}_{n - 1}}}$ (\emph{Step 1}) and the matrix ${{{\bf{\bar G}}_{n}}}$ (\emph{Step 3}) involving the matrix inversion of large size, which leads to high computational complexity as well as high hardware complexity. To this end, next we will propose a low-complexity algorithm to obtain ${{\bf{\bar p}}_n^{{\rm{opt}}}}$ to avoid the complicated SVD and matrix inversion.

\subsection{Low-complexity algorithm to obtain the optimal solution}\label{S3.3}
We start by considering how to avoid the SVD involving high computational complexity as well as a large number of divisions, which are difficult to be implemented in hardware. We can observe from \emph{Step 1} that the SVD of ${{{\bf{\bar G}}_{n - 1}}}$ does not need to be computed to acquire ${{\bf{\Sigma }}}$ and ${{\bf{V}}}$, as only the first column ${{{\bf{v}}_1}}$ of ${{\bf{V}}}$ is enough to obtain ${{\bf{\bar p}}_n^{{\rm{opt}}}}$. This observation inspires us to exploit the simple power iteration algorithm~\cite{golub2012matrix}, which is used to compute the largest eigenvalue and the corresponding eigenvector of a diagonalizable matrix. Since ${{{\bf{\bar G}}_{n - 1}}}$ is a Hermitian matrix, it follows that: 1) ${{{\bf{\bar G}}_{n - 1}}}$ is also a diagonalizable matrix; 2) The singular values (right singular vectors) of ${{{\bf{\bar G}}_{n - 1}}}$ are same as the eigenvalues (eigenvectors). Therefore, the power iteration algorithm can be also utilized to compute ${{{\bf{v}}_1}}$ as well as the largest singular value ${{\Sigma _1}}$ of ${{{\bf{\bar G}}_{n - 1}}}$ with low complexity.

More specifically, as shown by the pseudo-code in \textbf{Algorithm 1}, the power iteration algorithm starts with an initial solution ${{{\bf{u}}^{\left( 0 \right)}} \in \mathbb{C}{^{M \times 1}}}$, which is usually set as ${{\left[ {1,1, \cdots ,1} \right]^T}}$ without loss of generality~\cite{golub2012matrix}. In each iteration, it first computes the auxiliary vector ${{{\bf{z}}^{\left( s \right)}} = {{\bf{\bar G}}_{n - 1}}{{\bf{u}}^{\left( {s - 1} \right)}}}$ (${s}$ is the number of iterations) and then extracts the element of ${{{\bf{z}}^{\left( s \right)}}}$ having the largest amplitude as ${{m^{\left( s \right)}}}$. After that, ${{{\bf{u}}^{\left( s \right)}}}$ is updated as ${{{\bf{u}}^{\left( s \right)}} = \frac{{{{\bf{z}}^{\left( s \right)}}}}{{{m^{\left( s \right)}}}}}$ for the next iteration. The power iteration algorithm will stop until the number of iterations reaches the predefined number ${S}$. Finally, ${{m^{\left( S \right)}}}$ and ${{{\bf{u}}^{\left( S \right)}}/{\left\| {{{\bf{u}}^{\left( S \right)}}} \right\|_2}}$  will be output as the largest singular value ${{\Sigma _1}}$ and the first right singular vector  ${{{\bf{v}}_1}}$ of ${{{\bf{\bar G}}_{n - 1}}}$, respectively.

\begin{algorithm}[tp]
\caption{Power iteration algorithm}
\KwIn{(1) ${{{\bf{\bar G}}_{n - 1}}}$;
\\\hspace*{+9.3mm} (2) Initial solution ${{{\bf{u}}^{\left( 0 \right)}}}$;
\\\hspace*{+9.3mm} (3) Maximum number of iteration ${S}$}

 \textbf{for} ${1 \le s \le S}$
 \\\hspace*{+4.5mm} 1) ${{{\bf{z}}^{\left( s \right)}} = {{\bf{\bar G}}_{n - 1}}{{\bf{u}}^{\left( {s - 1} \right)}}}$
 \\\hspace*{+4.5mm} 2) ${{m^{\left( s \right)}} = \mathop {\arg \max }\limits_{z_i^{\left( s \right)}} \left| {z_i^{\left( s \right)}} \right|}$
 \\\hspace*{+4.5mm} 3) \textbf{if} ${1 \le s \le 2}$
 \\\hspace*{+12.5mm} ${{n^{\left( s \right)}} = {m^{\left( s \right)}}}$
 \\\hspace*{+8.5mm}  \textbf{else}
 \\\hspace*{+12.5mm} ${{n^{\left( s \right)}} = \frac{{{m^{\left( s \right)}}{m^{\left( {s - 2} \right)}} - {{\left( {{m^{\left( {s - 1} \right)}}} \right)}^2}}}{{{m^{\left( s \right)}} - 2{m^{\left( {s - 1} \right)}} + {m^{\left( {s - 2} \right)}}}}}$
 \\\hspace*{+8.5mm} \textbf{end if}
 \\\hspace*{+4.5mm} 4) ${{{\bf{u}}^{\left( s \right)}} = \frac{{{{\bf{z}}^{\left( s \right)}}}}{{{n^{\left( s \right)}}}}}$
 \\\textbf{end for}

\KwOut{(1) The largest singular value ${{\Sigma _1} = {n^{\left( S \right)}}}$
\\\hspace*{+12mm} (2) The first singular vector ${{{\bf{v}}_1} = \frac{{{{\bf{u}}^{\left( S \right)}}}}{{{{\left\| {{{\bf{u}}^{\left( S \right)}}} \right\|}_2}}}}$}
\end{algorithm}

According to~\cite{golub2012matrix}, we know that
\begin{equation}\label{eq18}
{m^{\left( s \right)}} = {\Sigma _1}\left[ {1 + {\cal O}\left( {{{\left( {\frac{{{\Sigma _2}}}{{{\Sigma _1}}}} \right)}^s}} \right)} \right],
\end{equation}
where ${{{\Sigma _2}}}$ is the second largest singular value of ${{{\bf{\bar G}}_{n - 1}}}$. From~(\ref{eq18}), we can conclude that ${{m^{\left( s \right)}}}$ will converges to ${{\Sigma _1}}$ as long as ${{\Sigma _1} \ne {\Sigma _2}}$. Similarly, when ${{\Sigma _1} \ne {\Sigma _2}}$, ${{{\bf{u}}^{\left( s \right)}}/{\left\| {{{\bf{u}}^{\left( s \right)}}} \right\|_2}}$ will also converge to ${{{\bf{v}}_1}}$, i.e.,
\begin{equation}\label{eq19}
\mathop {\lim }\limits_{s \to \infty } {m^{\left( s \right)}} = {\Sigma _1},\quad \mathop {\lim }\limits_{s \to \infty } \frac{{{{\bf{u}}^{\left( s \right)}}}}{{{{\left\| {{{\bf{u}}^{\left( s \right)}}} \right\|}_2}}} = {{\bf{v}}_1}.
\end{equation}

Although the power iteration algorithm is convergent, its convergence rate may be slow if ${{\Sigma _1} \approx {\Sigma _2}}$ based on~(\ref{eq18}). To solve this problem, we propose to utilize the Aitken acceleration method~\cite{bjorck2015numerical} to further increase the convergence rate of the power iteration algorithm. Specifically, we can compute
\begin{align}\label{eq20}
\left\{ \begin{array}{l}
{n^{\left( s \right)}} = {m^{\left( s \right)}}, \quad \quad \quad \quad \quad \quad \quad \mathrm{for}\;1 \le s \le 2,\\
{n^{\left( s \right)}} = \frac{{{m^{\left( s \right)}}{m^{\left( {s - 2} \right)}} - {{\left( {{m^{\left( {s - 1} \right)}}} \right)}^2}}}{{{m^{\left( s \right)}} - 2{m^{\left( {s - 1} \right)}} + {m^{\left( {s - 2} \right)}}}},\;\mathrm{for}\;2 < s \le S.
\end{array} \right.
\end{align}
Then, ${{{\bf{u}}^{\left( s \right)}}}$ and ${{\Sigma _1}}$ will be correspondingly changed to ${{{\bf{u}}^{\left( s \right)}} = \frac{{{{\bf{z}}^{\left( s \right)}}}}{{{n^{\left( s \right)}}}}}$ and ${{\Sigma _1} = {n^{\left( S \right)}}}$, respectively.

Next, we will focus on how to reduce the complexity to compute the matrices ${{{\bf{T}}_n}= {{\bf{I}}_N} + \frac{\rho }{{N{\sigma ^2}}}{\bf{H}}{{\bf{P}}_n}{\bf{P}}_n^H{{\bf{H}}^H}}$ and ${{{\bf{\bar G}}_n} = {\bf{R}}{{\bf{H}}^H}{\bf{T}}_n^{ - 1}{\bf{H}}{{\bf{R}}^H}}$, which involve the complicated matrix-to-matrix multiplication and matrix inversion of large size. In particular, with some standard mathematical manipulations, the computation of ${{{\bf{\bar G}}_n}}$ can be significantly simplified as shown by the following \textbf{Proposition 2}.

\vspace*{+2mm} \noindent\textbf{Proposition 2}. {\it The matrix ${{{\bf{\bar G}}_n} = {\bf{R}}{{\bf{H}}^H}{\bf{T}}_n^{ - 1}{\bf{H}}{{\bf{R}}^H}}$, where ${{{\bf{T}}_n}= {{\bf{I}}_N} + \frac{\rho }{{N{\sigma ^2}}}{\bf{H}}{{\bf{P}}_n}{\bf{P}}_n^H{{\bf{H}}^H}}$, can be simplified as
\begin{equation}\label{eq21}
{{\bf{\bar G}}_n} \approx {{\bf{\bar G}}_{n - 1}} - \frac{{\frac{\rho }{{N{\sigma ^2}}}\Sigma _1^2{{\bf{v}}_1}{\bf{v}}_1^H}}{{1 + \frac{\rho }{{N{\sigma ^2}}}{\Sigma _1}}},
\end{equation}
where ${{{\Sigma _1}}}$ and ${{{{\bf{v}}_1}}}$ are the largest singular value and first right singular vector of ${{{\bf{\bar G}}_{n - 1}}}$, respectively.}

\vspace*{+2mm}
\textit{Proof:} See Appendix B. \qed

\textbf{Proposition 2} implies that we can simply exploit ${{{\Sigma _1}}}$ and ${{{{\bf{v}}_1}}}$ that have been obtained by \textbf{Algorithm 1} as described above to update ${{{\bf{\bar G}}_n}}$, which only involves one vector-to-vector multiplication instead of the complicated matrix-to-matrix multiplication and matrix inversion. Note that the evaluation of computational complexity will be discussed in detail in Section III-E.

\subsection{Summary of the proposed SIC-based hybrid precoding}\label{S3.4}
Based on the discussion so far, the pseudo-code of the proposed SIC-based hybrid precoding can be summarized in \textbf{Algorithm 2}, which can be explained as follows. The proposed SIC-based hybrid precoding starts by computing the largest singular value ${{\Sigma _1}}$ and first right singular vector ${{{{\bf{v}}_1}}}$ of ${{{{\bf{\bar G}}}_{n-1}}}$, which is achieved by \textbf{Algorithm 1}. After that, according to Section III-B, the optimal precoding vector for the ${n}$th sub-antenna array can be obtained by utilizing ${{{{\bf{v}}_1}}}$. Finally, based on \textbf{Proposition 2}, ${{{{\bf{\bar G}}}_{n}}}$ can be updated with low complexity for the next iteration. This procedure will be executed until the last (${N}$th) sub-antenna array is considered. Finally, after ${N}$ iterations, the optimal digital, analog, and hybrid precoding matrix ${{\bf{D}}}$, ${{\bf{A}}}$, and ${{\bf{P}}}$ can be obtained, respectively.

\begin{algorithm}[h]
\caption{SIC-based hybrid precoding}
\KwIn{${{{\bf{\bar G}}_0}}$}

 \textbf{for} ${1 \le n \le N}$
 \\\hspace*{+4.5mm} 1) Compute ${{{{\bf{v}}_1}}}$ and ${{{\Sigma _1}}}$ of ${{{\bf{\bar G}}_{n-1}}}$ by \textbf{Algorithm 1}
 \\\hspace*{+4.5mm} 2) ${{\bf{\bar a}}_n^{{\rm{opt}}} = \frac{1}{{\sqrt M }}{e^{j{\rm{angle}}({{\bf{v}}_{\rm{1}}})}}}$, ${d_n^{{\rm{opt}}} = \frac{{{{\left\| {{{\bf{v}}_1}} \right\|}_1}}}{{\sqrt M }}}$,
 \\\hspace*{+9mm} ${{\bf{\bar p}}_n^{{\rm{opt}}} = \frac{1}{M}{\left\| {{{\bf{v}}_1}} \right\|_1}{e^{j{\rm{angle}}({{\bf{v}}_{\rm{1}}})}}}$ ~(\ref{eq14})-(\ref{eq16})
 \\\hspace*{+4.5mm} 3) ${{{\bf{\bar G}}_n}\! =\! {{\bf{\bar G}}_{n - 1}}\! -\! \frac{{\frac{\rho }{{N{\sigma ^2}}}\Sigma _1^2{{\bf{v}}_1}{\bf{v}}_1^H}}{{1 + \frac{\rho }{{N{\sigma ^2}}}{\Sigma _1}}}}$  (\textbf{Proposition 2})
 \\\textbf{end for}

\KwOut{(1) ${{\bf{D}} = {\rm{diag}}\left\{ {d_1^{{\rm{opt}}}, \cdots ,d_N^{{\rm{opt}}}} \right\}}$
\\\hspace*{+12mm} (2) ${{\bf{A}} = {\rm{diag}}\left\{ {{\bf{\bar a}}_1^{{\rm{opt}}}, \cdots ,{\bf{\bar a}}_N^{{\rm{opt}}}} \right\}}$
\\\hspace*{+12mm} (3) ${{\bf{P}} = {\bf{AD}}}$}
\end{algorithm}

It is worth pointing out that the idea of SIC-based hybrid precoding can be also extended to the combining at the user following the similar logic in~\cite{el2013spatially}. When the number of RF chains at the BS is smaller than that at the user, we first compute the optimal hybrid precoding matrix ${\bf{P}}$ according to \textbf{Algorithm 2}, where we assume that the combining matrix ${{\bf{Q}} = {\bf{I}}}$. Then, given the effective channel matrix ${{\bf{HP}}}$, we can similarly obtain the optimal hybrid combining matrix ${\bf{Q}}$ by referring to \textbf{Algorithm 2}, where the input ${{{\bf{\bar G}}_0}}$ and the optimal unconstrained solution ${{{{\bf{v}}_1}}}$ should be correspondingly replaced. Conversely,  when the number of RF chains at the BS is larger than that at the user, we can assume ${{\bf{P}} = {\bf{I}}}$ and obtain the optimal hybrid combining matrix ${\bf{Q}}$. After that, the optimal precoding matrix ${\bf{P}}$ can be acquired given the effective channel matrix ${{\bf{QH}}}$. Additionally, to further improve the performance, we can combine the above method with the ``Ping-pong" algorithm\cite{hur2013millimeter}, which involves an iteration procedure between the BS and the user, to jointly explore the optimal hybrid precoding and combining matrices pair. Further discussion about hybrid combining will be left for our future work.

\subsection{Complexity evaluation}\label{S3.5}
In this subsection, we provide the complexity evaluation of the proposed SIC-based hybrid precoding in terms of the required numbers of complex multiplications and divisions. From \textbf{Algorithm 2}, we can observe that the complexity of SIC-based hybrid precoding comes from the following four parts:

1) The first one originates from the computation of ${{{\bf{\bar G}}_0} = {\bf{R}}{{\bf{H}}^H}{\bf{H}}{{\bf{R}}^H}}$ according to~(\ref{eq11}). Note that ${\bf{R}}$ is a selection matrix and ${\bf{H}}$ has the size ${K \times NM}$. Therefore, this part involves ${KM^2}$ times of multiplications without any division.

2) The second one is from executing \textbf{Algorithm 1}. It can be observed that in each iteration we need to compute a matrix-to-vector multiplication ${{{\bf{z}}^{\left( s \right)}} = {{\bf{\bar G}}_{n - 1}}{{\bf{u}}^{\left( {s - 1} \right)}}}$ together with the Aitken acceleration method~(\ref{eq20}). Therefore, we totally require ${S\left( {{M^2} + 2} \right) - 4}$ and ${(2S - 2)}$ times of multiplications and divisions, respectively.

3) The third one stems from acquiring the optimal solution ${{\bf{\bar p}}_n^{{\rm{opt}}}}$ in step 2 of \textbf{Algorithm 2}. We can find that this part is quite simple, which only needs 2 times of multiplications without any division, since ${{{{\bf{v}}_1}}}$ has been obtained and ${\frac{1}{{\sqrt M }}}$ is a fixed constant.

4) The last one comes from the update of ${{{{\bf{\bar G}}}_n}}$. According to \textbf{Proposition 2}, we know that this part mainly involves a outer product ${{{{\bf{v}}_1}{\bf{v}}_1^H}}$. Thus, it requires ${M^2}$ times of multiplications with only one division.

To sum up, the proposed SIC-based hybrid precoding approximately requires ${{M^2}\left( {NS + K} \right)}$ times of multiplications and ${{2NS}}$ times of divisions. It is worth pointing out that the recently proposed spatially sparse precoding~\cite{el2013spatially} requires ${{\cal O}\left( {{N^4}M + {N^2}{L^2} + {N^2}{M^2}L} \right)}$ times of multiplications and ${{\cal O}\left( {2{N^3}} \right)}$ times of divisions, where ${L}$ is the number of effective channel paths as defined in~(\ref{eq3}). Considering the typical mmWave MIMO system with ${N=8}$, ${M=8}$, ${K=16}$, ${L=3}$~\cite{el2013spatially}, we can observe that the complexity of SIC-based hybrid precoding is about ${4 \times {10^3}}$ times of multiplications and ${{10^2}}$ times of divisions, where we set ${S=5}$ that is enough to guarantee the performance as will be verified in Section IV. By contrast, the complexity of the spatially sparse precoding is about ${5 \times {10^4}}$ times of multiplications and ${10^3}$ times of divisions. Therefore, the proposed SIC-based hybrid precoding enjoys much lower complexity, which is only about 10\% as complex as that of the spatially sparse precoding.

\section{Simulation Results}\label{S4}
In this section, we provide the simulation results of the achievable rate to evaluate the performance of the proposed SIC-based hybrid precoding. We compare the performance of SIC-based hybrid precoding with the recently proposed spatially sparse precoding~\cite{el2013spatially} and the optimal unconstrained precoding based on the SVD of the channel matrix, which are both with fully-connected architecture. Additionally, we also include the conventional analog precoding and the optimal unconstrained precoding  (i.e., ${{\bf{\bar p}}_n^{{\rm{opt}}} = {{\bf{v}}_1}}$) which are both with sub-connected architecture~\cite{el2013multimode} as benchmarks for comparison.

The simulation parameters are described as follows. We generate the channel matrix according to the channel model~\cite{alkhateeb2014channel} described in Section~\ref{S2}.  The number of effective channel paths is ${L = 3}$. The carrier frequency is set as 28GHz~\cite{roh2014millimeter}. Both the transmit and receive antenna arrays are ULAs with antenna spacing ${d = \lambda /2}$. Since the BS usually employs the directional antennas to eliminate interference and increase antenna gain~\cite{pi2011introduction}, the AoDs are assumed to follow the uniform distribution within ${\left[ { - \frac{\pi }{6},\frac{\pi }{6}} \right]}$. Meanwhile, due to the random position of users, we assume that the AoAs follow the uniform distribution within ${\left[ { - \pi ,\pi } \right]}$, which means the omni-directional antennas are adopted by users. Furthermore, we set the maximum number of iterations ${S=5}$ to run \textbf{Algorithm 2}. Finally, the signal-to-noise ratio (SNR) is defined as ${\frac{\rho }{{{\sigma ^2}}}}$.

\begin{figure}[tp]
\begin{center}
\hspace*{+1mm}\includegraphics[width=0.95\linewidth]{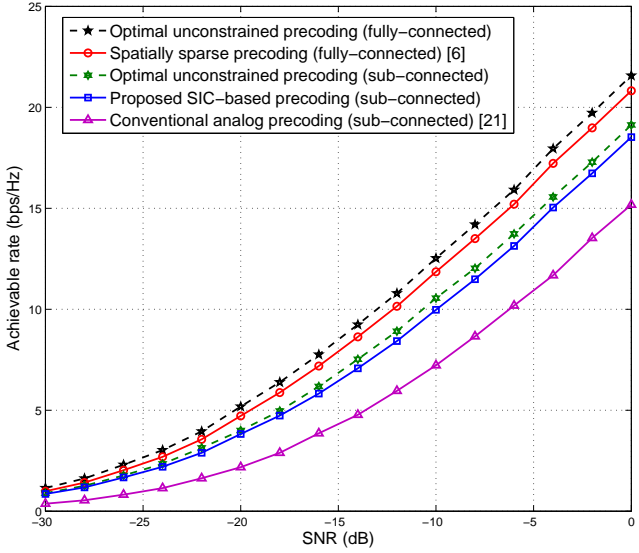}
\end{center}
\vspace*{-3mm}\caption{Achievable rate comparison for an ${NM \times K = 64 \times 16}$ (${N = 8}$) mmWave MIMO system.} \label{FIG3}
\vspace*{+2mm}\end{figure}

\begin{figure}[tp]
\begin{center}
\hspace*{+1mm}\includegraphics[width=0.95\linewidth]{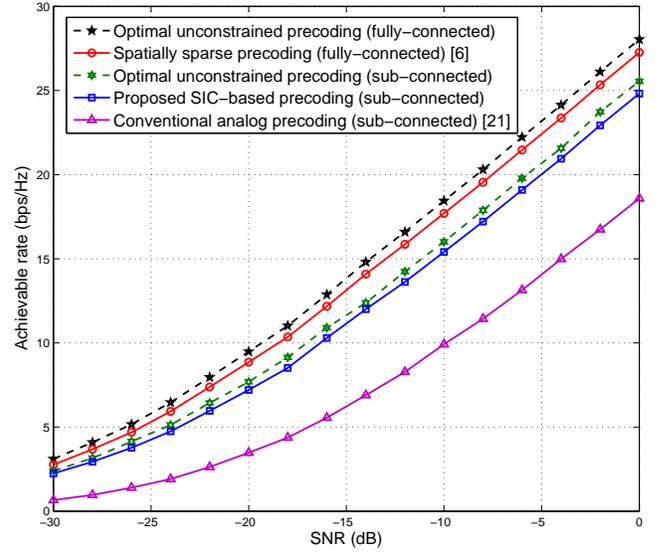}
\end{center}
\vspace*{-3mm}\caption{Achievable rate comparison for an ${NM \times K = 128 \times 32}$ (${N = 16}$) mmWave MIMO system.} \label{FIG4}
\vspace*{+2mm}\end{figure}

Fig. 3 shows the achievable rate comparison in mmWave MIMO system, where ${NM \times K = 64 \times 16}$ and the number of RF chains is ${N=8}$. We can observe from Fig. 3 that the proposed SIC-based hybrid precoding outperforms the conventional analog precoding with sub-connected architecture  in whole simulated SNR range. Meanwhile, Fig. 3 also verifies the near-optimal performance of SIC-based hybrid precoding, since it can achieve about 99\% of the rate achieved by the optimal unconstrained precoding with sub-connected architecture.

Fig. 4 compares the achievable rate in mmWave MIMO system with ${NM \times K = 128 \times 32}$ and ${N=16}$, where we can observe similar trends as those from Fig. 3. More importantly, Fig. 3 and Fig. 4  show that the performance of SIC-based hybrid precoding is also close to the spatially sparse precoding and the optimal unconstrained precoding with fully-connected architecture. For example, when SNR = 0 dB, our method can achieve more than 90\% of the rate achieved by the near-optimal spatially sparse precoding in both simulated mmWave MIMO configurations. Considering the low energy consumption and computational complexity of the proposed SIC-based hybrid precoding as analyzed before, we can further conclude that SIC-based hybrid precoding can achieve much better trade-off among the performance, energy consumption, and computational complexity.

Fig. 5 provides a achievable rate comparison in  mmWave MIMO systems against the numbers of BS and user antennas, where ${NM =K}$ and the number of RF chains is fixed to ${N=8}$. We can find that the performance of the proposed SIC-based hybrid precoding can be improved by increasing the number of BS and user antennas, which involves much lower energy consumption than increasing the number of energy-intensive RF chains~\cite{balanis2012antenna}.

Fig. 6 shows the achievable rate comparison against the numbers of user antennas ${K}$, where ${NM=64}$ and ${N=8}$. We can imply from Fig. 6 that the performance loss of SIC-based hybrid precoding due to the sub-connected architecture can be compensated by increasing the number of user antennas ${K}$. For example, the achievable rate of SIC-based hybrid precoding when ${K = 30}$ is the same as that of the spatially sparse precoding when ${K=20}$. Note that in this case, the required number of phase shifters of SIC-based hybrid precoding is ${NM=64}$ and each RF chain only needs to drive 8 BS antennas, while for the spatially sparse precoding, the number of required phase shifters  is ${N^2M=512}$ and each RF chain has to drive 64 BS antennas. By contrast, the cost of increasing the number of user antennas ${K}$ will be negligible since the power consumption of user antenna is usually small~\cite{balanis2012antenna}. Therefore, we can conclude that the proposed SIC-based hybrid precoding is more energy-efficient and implementation-practical.

\begin{figure}[tp]
\begin{center}
\hspace*{+1mm}\includegraphics[width=0.95\linewidth]{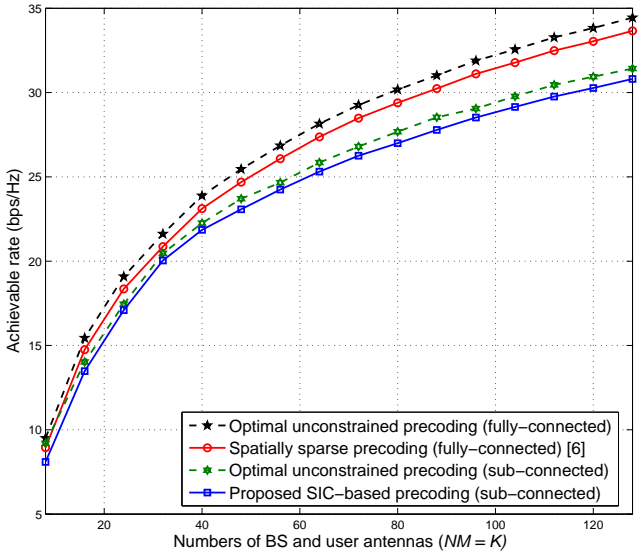}
\end{center}
\vspace*{-3mm}\caption{Achievable rate comparison against the numbers of BS and user antennas (${NM =K}$), where ${N=8}$.} \label{FIG4}
\vspace*{+2mm}\end{figure}

\begin{figure}[tp]
\begin{center}
\hspace*{+1mm}\includegraphics[width=0.95\linewidth]{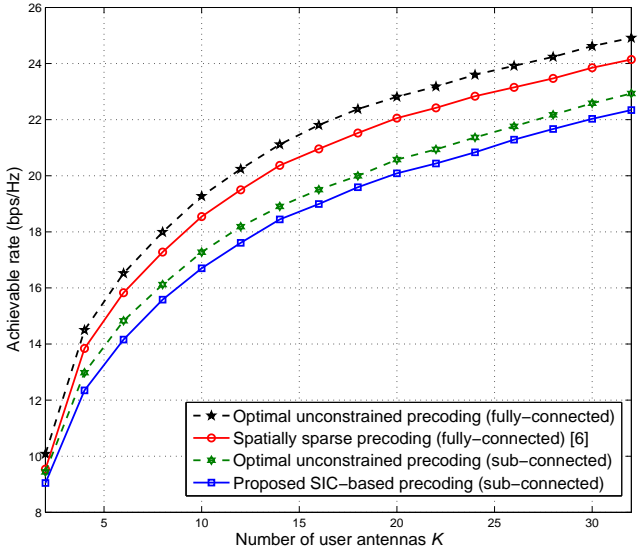}
\end{center}
\vspace*{-2mm}\caption{Achievable rate comparison against the number of user antennas ${K}$, where ${NM=64}$ and ${N=8}$.} \label{FIG4}
\vspace*{+2mm}\end{figure}

\section{Conclusions}\label{S5}
In this paper, we proposed a SIC-based hybrid precoding with sub-connected architecture for mmWave MIMO systems. We first showed that the total achievable rate optimization problem with non-convex constraints can be decomposed into a series of sub-rate optimization problems, each of which only considers one sub-antenna array. Then, we proved that the sub-rate optimization problem of each sub-antenna array can be solved by simply seeking a precoding vector sufficiently close to the unconstrained optimal solution. Finally, a low-complexity algorithm was proposed to realize SIC-based precoding without the complicated SVD and  matrix inversion. Complexity evaluation showed that the complexity of the proposed SIC-based hybrid precoding is only about 10\% as complex as that ot the recently proposed spatially sparse precoding with fully-connected architecture in typical mmWave MIMO system. Simulation results verified the near-optimal performance of SIC-based hybrid precoding, and implied that the performance loss induced by sub-connected architecture can be compensated by increasing the number of antennas. This may be a reasonable tradeoff versus increasing the number of phase shifters as required in the fully-connected architecture.
Our further work will focus on the limited feedback scenario, where the channel state information may be not perfect and the angles of phase shifters are quantified.

\balance

\vspace*{+3mm}
\section*{Appendix A\\ Proof of Proposition 1}
Define the target of the optimization problem~(\ref{eq10}) as
\begin{equation}\label{eq51}
{R_n} = {\log _2}\left( {1 + \frac{\rho }{{N{\sigma ^2}}}{\bf{\bar p}}_n^H{{{\bf{\bar G}}}_{n - 1}}{{{\bf{\bar p}}}_n}} \right),
\end{equation}
and the SVD of ${{{{{\bf{\bar G}}}_{n - 1}}}}$ as ${{{\bf{\bar G}}_{n - 1}} = {\bf{V\Sigma }}{{\bf{V}}^H}}$. Then, by separating  the matrices ${{\bf{\Sigma }}}$ and ${{\bf{V}}}$ into two parts:
\begin{equation}\label{eq52}
{\bf{\Sigma }} = \left[ {\begin{array}{*{20}{c}}
{{\Sigma _1}}&0\\
0&{{{\bf{\Sigma }}_2}}
\end{array}} \right],\quad {\bf{V}} = \left[ {{{\bf{v}}_1}\;{{\bf{V}}_2}} \right],
\end{equation}
${R_n}$ in~(\ref{eq51}) can be rewritten as
\begin{align}\label{eq53}
{R_n}& = {\log _2}\left( {1 + \frac{\rho }{{N{\sigma ^2}}}{\bf{\bar p}}_n^H{{{\bf{\bar G}}}_{n - 1}}{{{\bf{\bar p}}}_n}} \right) \nonumber \\
 &= {\log _2}\left( {1 + \frac{\rho }{{N{\sigma ^2}}}{\bf{\bar p}}_n^H{\bf{V\Sigma }}{{\bf{V}}^H}{{{\bf{\bar p}}}_n}} \right) \nonumber \\
 &= {\log _2}\left( {1 + \frac{\rho }{{N{\sigma ^2}}}} \right. \nonumber \\
&\left. {\quad \quad \quad \; \; \times {\bf{\bar p}}_n^H\left[ {{{\bf{v}}_1}\;{{\bf{V}}_2}} \right]\left[ {\begin{array}{*{20}{c}}
{{\Sigma _1}}&0\\
0&{{{\bf{\Sigma }}_2}}
\end{array}} \right]{{\left[ {{{\bf{v}}_1}\;{{\bf{V}}_2}} \right]}^H}{{{\bf{\bar p}}}_n}} \right) \nonumber \\
& = {\log _2}\left( {1 + \frac{\rho }{{N{\sigma ^2}}}{\bf{\bar p}}_n^H{{\bf{v}}_1}{\Sigma _1}{\bf{v}}_1^H{{{\bf{\bar p}}}_n}} \right. \nonumber \\
&\left. {\quad \quad \quad \; \; + \frac{\rho }{{N{\sigma ^2}}}{\bf{\bar p}}_n^H{{\bf{V}}_2}{{\bf{\Sigma }}_2}{\bf{V}}_2^H{{{\bf{\bar p}}}_n}} \right).
\end{align}

Since we aim to find a vector ${{{{{\bf{\bar p}}}_n}}}$ sufficiently ``close" to ${{{{\bf{v}}_1}}}$, it is reasonable to assume that ${{{{{\bf{\bar p}}}_n}}}$ is approximately orthogonal to the matrix ${{{{\bf{V}}_2}}}$, i.e., ${{\bf{\bar p}}_n^H{{\bf{V}}_2} \approx {\bf{0}}}$~\cite{el2013spatially}. Then,~(\ref{eq53}) can be simplified as
\begin{align}\label{eq54}
{R_n} &\approx {\log _2}\left( {1 + \frac{{\rho {\Sigma _1}}}{{N{\sigma ^2}}}{\bf{\bar p}}_n^H{{\bf{v}}_1}{\bf{v}}_1^H{{{\bf{\bar p}}}_n}} \right) \nonumber \\
&\mathop = \limits^{\left( a \right)} {\log _2}\left( {1 + \frac{{\rho {\Sigma _1}}}{{N{\sigma ^2}}}} \right) \nonumber \\
&\quad  + {\log _2}\left( {1 - {{\left( {1 + \frac{{\rho {\Sigma _1}}}{{N{\sigma ^2}}}} \right)}^{ - 1}}\frac{{\rho {\Sigma _1}}}{{N{\sigma ^2}}}\left( {1\! -\! {\bf{\bar p}}_n^H{{\bf{v}}_1}{\bf{v}}_1^H{{{\bf{\bar p}}}_n}} \right)} \right)  \nonumber \\
&\mathop  \approx \limits^{\left( b \right)} {\log _2}\left( {1 + \frac{{\rho {\Sigma _1}}}{{N{\sigma ^2}}}} \right) + {\log _2}\left( {{\bf{\bar p}}_n^H{{\bf{v}}_1}{\bf{v}}_1^H{{{\bf{\bar p}}}_n}} \right)
\end{align}
where ${{\left( a \right)}}$ is obtained by using the formula ${{\bf{I}}+ {\bf{XY}}= \left( {{\bf{I}}+ {\bf{X}}} \right)\left( {{\bf{I}} - {{\left( {{\bf{I}} + {\bf{X}}} \right)}^{ - 1}}{\bf{X}}\left( {{\bf{I}} - {\bf{Y}}} \right)} \right)}$~\cite{el2013spatially}, where we define ${{\bf{X}}= \frac{{\rho {\Sigma _1}}}{{N{\sigma ^2}}}}$ and ${{\bf{Y}} = {\bf{\bar p}}_n^H{{\bf{v}}_1}{\bf{v}}_1^H{{\bf{\bar p}}_n}}$; ${{\left( b \right)}}$ is valid by employing the high SNR approximation~\cite{tse2005fundamentals}, i.e., \begin{align}\label{eq55}
{\left( {1 + \frac{{\rho {\Sigma _1}}}{{N{\sigma ^2}}}} \right)^{ - 1}}\frac{{\rho {\Sigma _1}}}{{N{\sigma ^2}}} \approx 1.
\end{align}

From~(\ref{eq54}), we can observe that maximizing ${R_n}$ is equivalent to maximizing ${{\bf{\bar p}}_n^H{{\bf{v}}_1}{\bf{v}}_1^H{{\bf{\bar p}}_n} = \left\| {{\bf{\bar p}}_n^H{{\bf{v}}_1}} \right\|_2^2}$, the square of inner product between two vectors ${{{\bf{\bar p}}_n}}$ and ${{{{\bf{v}}_1}}}$. Note that ${{{{\bf{v}}_1}}}$ is a fixed vector. Therefore, exploring a vector ${{{\bf{\bar p}}_n}}$, which has the largest projection on ${{{{\bf{v}}_1}}}$, will lead to the smallest Euclidean distance to ${{{{\bf{v}}_1}}}$ as well. Based on this fact, we can conclude that the optimization problem~(\ref{eq10}) is equivalent to the following problem
\begin{equation}\label{eq55}
{\bf{\bar p}}_n^{{\rm{opt}}} = \mathop {\arg \min }\limits_{{{{\bf{\bar p}}}_n} \in \bar {\cal F}} \left\| {{{\bf{v}}_1} - {{{\bf{\bar p}}}_n}} \right\|_2^2.
\end{equation}
\qed

\vspace*{+6mm}
\section*{Appendix B\\ Proof of Proposition 2}
We first consider the matrix ${{{\bf{T}}_n} = {{\bf{I}}_N} + \frac{\rho }{{N{\sigma ^2}}}{\bf{H}}{{\bf{P}}_n}{\bf{P}}_n^H{{\bf{H}}^H}}$, which should be inversed to compute ${{{{\bf{\bar G}}}_n}}$~(\ref{eq11}). By partitioning ${{{\bf{P}}_n}}$ as ${{{\bf{P}}_n} = \left[ {{{\bf{P}}_{n - 1}}\;{{\bf{p}}_n}} \right]}$, ${{{\bf{T}}_n}}$ can be rewritten as
\begin{align}\label{eq56}
{{\bf{T}}_n} &= {{\bf{I}}_N} + \frac{\rho }{{N{\sigma ^2}}}{\bf{H}}{{\bf{P}}_n}{\bf{P}}_n^H{{\bf{H}}^H} \nonumber \\
&= {{\bf{I}}_N} + \frac{\rho }{{N{\sigma ^2}}}{\bf{H}}\left[ {{{\bf{P}}_{n - 1}}\;{{\bf{p}}_n}} \right]{\left[ {{{\bf{P}}_{n - 1}}\;{{\bf{p}}_n}} \right]^H}{{\bf{H}}^H} \nonumber \\
&= {{\bf{I}}_N} + \frac{\rho }{{N{\sigma ^2}}}{\bf{H}}{{\bf{P}}_{n- 1}}{\bf{P}}_{n- 1}^H{{\bf{H}}^H} + \frac{\rho }{{N{\sigma ^2}}}{\bf{H}}{{\bf{p}}_n}{\bf{p}}_n^H{{\bf{H}}^H} \nonumber \\
&= {{\bf{T}}_{n - 1}} + \frac{\rho }{{N{\sigma ^2}}}{\bf{H}}{{\bf{p}}_n}{\bf{p}}_n^H{{\bf{H}}^H}.
\end{align}
Then, by utilizing the Sherman-Morrison formula~\cite[Eq 2.1.4]{golub2012matrix}
\begin{equation}\label{eq71}
{\left( {{\bf{A}} + {\bf{u}}{{\bf{v}}^T}} \right)^{ - 1}} = {{\bf{A}}^{ - 1}} - \frac{{{{\bf{A}}^{ - 1}}{\bf{u}}{{\bf{v}}^T}{{\bf{A}}^{ - 1}}}}{{1 + {{\bf{v}}^T}{{\bf{A}}^{ - 1}}{\bf{u}}}},
\end{equation}
${{\bf{T}}_n^{ - 1}}$ can be presented as
\begin{align}\label{eq57}
{\bf{T}}_n^{ - 1} &= {\left( {{{\bf{T}}_{n - 1}} + \frac{\rho }{{N{\sigma ^2}}}{\bf{H}}{{\bf{p}}_n}{\bf{p}}_n^H{{\bf{H}}^H}} \right)^{- 1}} \nonumber \\
 &= {\bf{T}}_{n - 1}^{ - 1} - \frac{{\frac{\rho }{{N{\sigma ^2}}}{\bf{T}}_{n - 1}^{ - 1}{\bf{H}}{{\bf{p}}_n}{\bf{p}}_n^H{{\bf{H}}^H}{\bf{T}}_{n - 1}^{ - 1}}}{{1 + \frac{\rho }{{N{\sigma ^2}}}{\bf{p}}_n^H{{\bf{H}}^H}{\bf{T}}_{n - 1}^{ - 1}{\bf{H}}{{\bf{p}}_n}}}.
\end{align}
Substituting~(\ref{eq57}) into ${{{\bf{G}}_n}={{\bf{H}}^H}{\bf{T}}_n^{ - 1}{\bf{H}}}$, we have
\begin{align}\label{eq58}
{{\bf{G}}_n} &={{\bf{H}}^H}{\bf{T}}_n^{ - 1}{\bf{H}} \nonumber \\
&={{\bf{H}}^H}\left( {{\bf{T}}_{n - 1}^{ - 1} - \frac{{\frac{\rho }{{{\sigma ^2}}}{\bf{T}}_{n - 1}^{ - 1}{\bf{H}}{{\bf{p}}_n}{\bf{p}}_n^H{{\bf{H}}^H}{\bf{T}}_{n - 1}^{ - 1}}}{{1 + \frac{\rho }{{{\sigma ^2}}}{\bf{p}}_n^H{{\bf{H}}^H}{\bf{T}}_{n - 1}^{ - 1}{\bf{H}}{{\bf{p}}_n}}}} \right){\bf{H}} \nonumber \\
& ={{\bf{G}}_{n -1}} - \frac{{\frac{\rho }{{{\sigma ^2}}}{{\bf{G}}_{n - 1}}{{\bf{p}}_n}{\bf{p}}_n^H{{\bf{G}}_{n - 1}}}}{{1 + \frac{\rho }{{{\sigma ^2}}}{\bf{p}}_n^H{{\bf{G}}_{n - 1}}{{\bf{p}}_n}}}.
\end{align}
Then, according to~(\ref{eq11}), ${{{{\bf{\bar G}}}_n}}$ can be obtained by
\begin{align}\label{eq59}
{{{\bf{\bar G}}}_n} &= {\bf{R}}{{\bf{G}}_{n}}{{\bf{R}}^H} \nonumber \\
 &= {\bf{R}}\left( {{{\bf{G}}_{n - 1}} - \frac{{\frac{\rho }{{N{\sigma ^2}}}{{\bf{G}}_{n - 1}}{{\bf{p}}_n}{\bf{p}}_n^H{{\bf{G}}_{n - 1}}}}{{1 + \frac{\rho }{{N{\sigma ^2}}}{\bf{p}}_n^H{{\bf{G}}_{n-1}}{{\bf{p}}_n}}}} \right){{\bf{R}}^H} \nonumber \\
 &= {{{\bf{\bar G}}}_{n -1}} - \frac{{\frac{\rho }{{N{\sigma ^2}}}{{{\bf{\bar G}}}_{n - 1}}{{{\bf{\bar p}}}_n}{\bf{\bar p}}_n^H{{{\bf{\bar G}}}_{n - 1}}}}{{1 + \frac{\rho }{{N{\sigma ^2}}}{\bf{\bar p}}_n^H{{{\bf{\bar G}}}_{n- 1}}{{{\bf{\bar p}}}_n}}}.
\end{align}
Note that in Section III-B, we have obtained the optimal solution ${{\bf{\bar p}}_n^{{\rm{opt}}}}$ which is sufficiently close to ${{{{\bf{v}}_1}}}$. Thus,~(\ref{eq59}) can be well approximated by replacing ${{{{{\bf{\bar p}}}_n}}}$ with ${{{{\bf{v}}_1}}}$ as
\begin{align}\label{eq60}
{{{\bf{\bar G}}}_n} &= {{{\bf{\bar G}}}_{n-1}} - \frac{{\frac{\rho }{{N{\sigma ^2}}}{{{\bf{\bar G}}}_{n-1}}{{{\bf{\bar p}}}_n}{\bf{\bar p}}_n^H{{{\bf{\bar G}}}_{n-1}}}}{{1 + \frac{\rho }{{N{\sigma ^2}}}{\bf{\bar p}}_n^H{{{\bf{\bar G}}}_{n-1}}{{{\bf{\bar p}}}_n}}} \nonumber \\
 &\approx {{{\bf{\bar G}}}_{n-1}} - \frac{{\frac{\rho }{{N{\sigma ^2}}}{{{\bf{\bar G}}}_{n-1}}{{\bf{v}}_1}{\bf{v}}_1^H{{{\bf{\bar G}}}_{n-1}}}}{{1 + \frac{\rho }{{N{\sigma ^2}}}{\bf{v}}_1^H{{{\bf{\bar G}}}_{n - 1}}{{\bf{v}}_1}}} \nonumber \\
&\mathop  = \limits^{\left( a \right)} {{{\bf{\bar G}}}_{n - 1}} - \frac{{\frac{\rho }{{N{\sigma ^2}}}\Sigma _1^2{{\bf{v}}_1}{\bf{v}}_1^H}}{{1 + \frac{\rho }{{N{\sigma ^2}}}{\Sigma _1}}},
\end{align}
where ${{\left( a \right)}}$ is true due to fact that ${{\bf{v}}_1^H{{\bf{\bar G}}_{n - 1}}= {\Sigma _1}{\bf{v}}_1^H}$, since ${{{\bf{\bar G}}_{n - 1}}}$ is an Hermitian matrix.
\qed

\bibliography{IEEEabrv,Gao1Ref}

\vspace*{-10mm}
\begin{IEEEbiography}[{\includegraphics[width=1in,height=1.25in,clip,keepaspectratio]{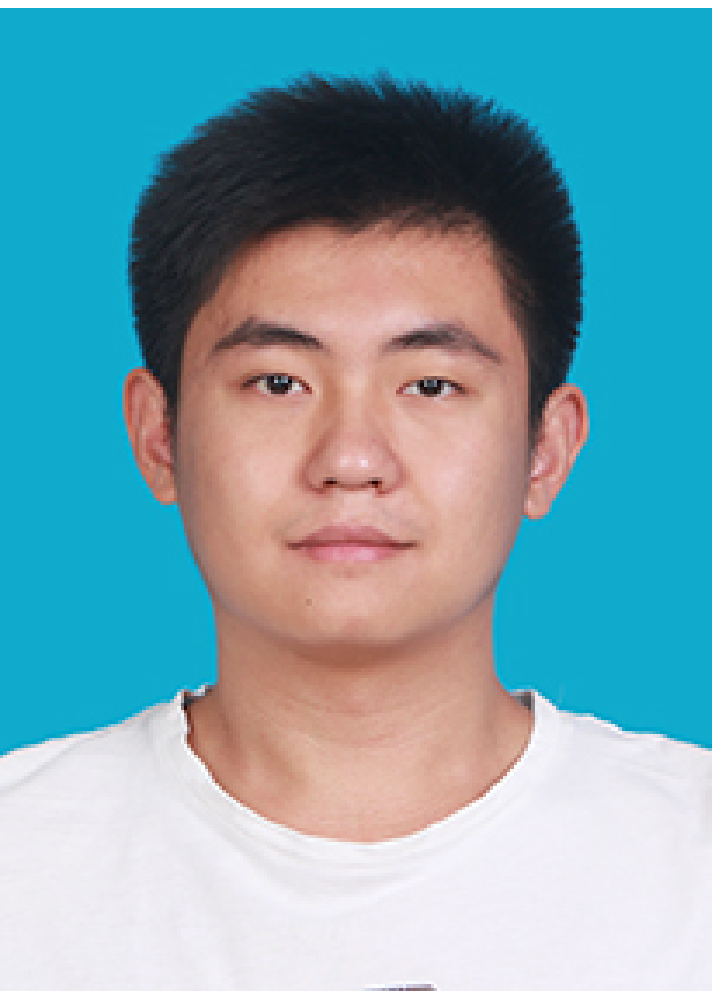}}]
{Xinyu Gao}(S'14) received the B.E. degree of Communication
Engineering from Harbin Institute of Technology, Heilongjiang, China in 2014. He is currently working towards Ph. D. degree in Electronic Engineering from Tsinghua University, Beijing, China. His research interests include massive MIMO and mmWave communications, with the emphasis on signal detection and precoding. He has published several journal and conference papers in IEEE Journal on Selected Areas in Communications, IEEE Transaction on Vehicular Technology, IEEE ICC, IEEE GLOBECOM, etc.
\end{IEEEbiography}

\vspace*{-10mm}
\begin{IEEEbiography}[{\includegraphics[width=1in,height=1.25in,clip,keepaspectratio]{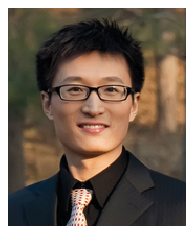}}]
{Linglong Dai}(M'11- SM'14) received the B.S. degree from Zhejiang University
in 2003, the M.S. degree (with the highest honors) from the China Academy
of Telecommunications Technology (CATT) in 2006, and the Ph.D. degree (with
the highest honors) from Tsinghua University, Beijing, China, in 2011. From
2011 to 2013, he was a Postdoctoral Fellow at the Department of Electronic
Engineering, Tsinghua University, and then since July 2013, became an Assistant
Professor with the same Department. His research interests are in wireless
communications with the emphasis on OFDM, MIMO, synchronization, channel
estimation, multiple access techniques, and wireless positioning. He has
published over 50 journal and conference papers. He has received IEEE Scott Helt Memorial Award in 2015 (IEEE Transactions on Broadcasting Best Paper Award), IEEE ICC
Best Paper Award in 2014, URSI Young Scientists Award in 2014, National Excellent
Doctoral Dissertation Nomination Award in 2013, IEEE ICC Best Paper
Award in 2013, Excellent Doctoral Dissertation of Beijing in 2012, Outstanding
Ph.D. Graduate of Tsinghua University in 2011.
\end{IEEEbiography}

\vspace*{-10mm}
\begin{IEEEbiography}[{\includegraphics[width=1in,height=1.25in,clip,keepaspectratio]{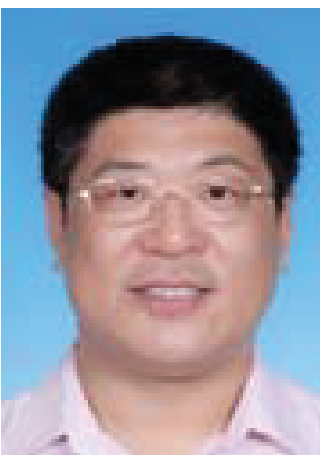}}]
{Shuangfeng Han} received his M.S. and Ph.D. degrees in electrical engineering from Tsinghua University in 2002 and 2006 respectively. He joined Samsung Electronics as a senior engineer in 2006 working on MIMO, MultiBS MIMO etc. From 2012, he is a senior project manager in the Green Communication Research Center at the China Mobile Research Institute. His research interests are green 5G, massive MIMO, full duplex, NOMA and EE-SE co-design.
\end{IEEEbiography}

\begin{IEEEbiography}[{\includegraphics[width=1in,height=1.25in,clip,keepaspectratio]{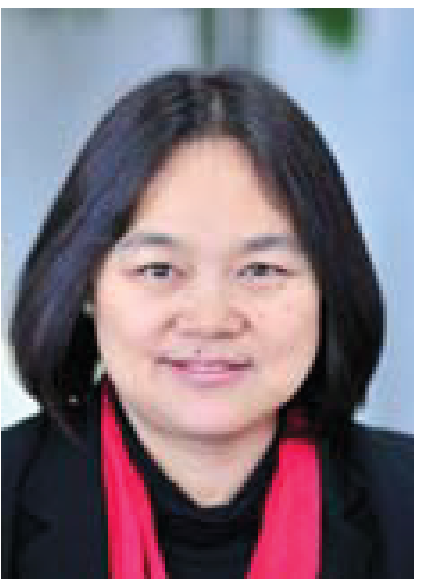}}]
{Chih-Lin I} received her Ph.D. degree in electrical engineering from Stanford University. She has been working at multiple world-class companies and research institutes leading the R${{\rm{\& }}}$D, including AT${{\rm{\& }}}$T Bell Labs; AT${{\rm{\& }}}$T HQ, ITRI of Taiwan, and ASTRI of Hong Kong. She received the IEEE Trans. on Commu. Stephen Rice Best Paper Award and is a winner of the CCCP National 1000 Talent program. Currently, she is China Mobile's chief scientist of wireless technologies and has established the Green Communications Research Center, spearheading major initiatives including key 5G technology R${{\rm{\& }}}$D; high energy efficiency system architectures, technologies and devices; green energy; and C-RAN and soft base stations. She was an elected Board Member of IEEE ComSoc, Chair of the ComSoc Meetings and Conferences Board, and Founding Chair of the IEEE WCNC Steering Committee. She is currently an Executive Board Member of GreenTouch and a Network Operator Council Member of ETSI NFV. Her research interests are green communications, C-RAN, network convergence, bandwidth refarming, EE-SE co-design, massive MIMO, and active antenna arrays.
\end{IEEEbiography}

\begin{IEEEbiography}[{\includegraphics[width=1in,height=1.25in,clip,keepaspectratio]{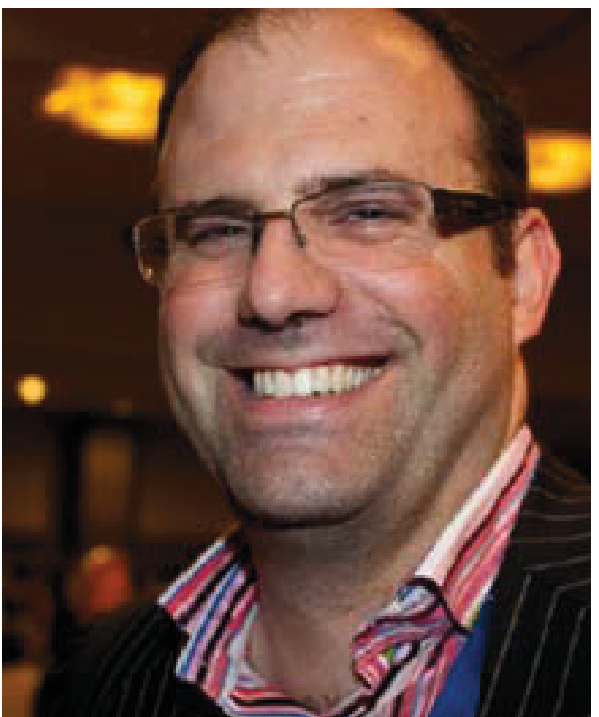}}]
{Robert W. Heath, Jr.} (S'96- M'01- SM'06- F'11) received the B.S. and M.S. degrees from the University of Virginia, Charlottesville, VA, in 1996 and 1997 respectively, and the Ph.D. from Stanford University, Stanford, CA, in 2002, all in electrical engineering. From 1998 to 2001, he was a Senior Member of the Technical Staff then a Senior Consultant
at Iospan Wireless Inc, San Jose, CA where he worked on the design and implementation of the physical and link layers of the first commercial MIMO-OFDM communication system. Since January 2002, he has been with the Department of Electrical and Computer Engineering at The University of Texas at Austin where he is a Cullen Trust for Higher Education Endowed Professor, and is Director of the Wireless Networking and Communications Group. He is also President and CEO of MIMO Wireless Inc. and Chief Innovation Officer at Kuma Signals LLC. His research interests include several aspects of wireless communication and signal processing: limited feedback techniques, multihop networking, multiuser and multicell MIMO, interference alignment, adaptive video transmission, manifold signal processing, and millimeter wave communication techniques.

Dr. Heath has been an Editor for the IEEE TRANSACTIONS ON COMMUNICATION, an Associate Editor for the IEEE TRANSACTIONS ON VEHICULAR
TECHNOLOGY, lead guest editor for an IEEE JOURNAL ON SELECTED AREAS IN COMMUNICATIONS special issue on limited feedback communication, and lead guest editor for an IEEE JOURNAL ON SELECTED TOPICS IN SIGNAL PROCESSING special issue on Heterogenous Networks. He currently serves on the steering committee for the IEEE TRANSACTIONS ON WIRELESS COMMUNICATIONS. He was a member of the Signal Processing for Communications Technical Committee in the IEEE Signal Processing Society
and is a former Chair of the IEEE COMSOC Communications Technical
Theory Committee. He was a technical co-chair for the 2007 Fall Vehicular
Technology Conference, general chair of the 2008 Communication Theory
Workshop, general co-chair, technical co-chair and co-organizer of the 2009
IEEE Signal Processing for Wireless Communications Workshop, local coorganizer
for the 2009 IEEE CAMSAP Conference, technical co-chair for
the 2010 IEEE International Symposium on Information Theory, the technical
chair for the 2011 Asilomar Conference on Signals, Systems, and Computers,
general chair for the 2013 Asilomar Conference on Signals, Systems, and
Computers, founding general co-chair for the 2013 IEEE GlobalSIP conference,
and is technical co-chair for the 2014 IEEE GLOBECOM conference.
Dr. Heath was a co-author of best student paper awards at IEEE VTC 2006
Spring, WPMC 2006, IEEE GLOBECOM 2006, IEEE VTC 2007 Spring,
and IEEE RWS 2009, as well as co-recipient of the Grand Prize in the 2008
WinTech WinCool Demo Contest. He was co-recipient of the 2010 and 2013
EURASIP Journal on Wireless Communications and Networking best paper
awards and the 2012 Signal Processing Magazine best paper award. He was
a 2003 Frontiers in Education New Faculty Fellow. He is also a licensed
Amateur Radio Operator and is a registered Professional Engineer in Texas.
\end{IEEEbiography}

\end{document}